\documentclass[]{aastex631}
\usepackage[caption=false]{subfig}
\usepackage{multirow}
\usepackage{graphicx}
\usepackage{booktabs}
\usepackage{amsmath,amssymb}
\usepackage[T1]{fontenc}
\usepackage{threeparttable}

\shortauthors{Cheng et al.}

\begin{document}
\title{TeV and keV-MeV Excesses as Probes for Hadronic Process in BL Lacertaes}

\correspondingauthor{Xiao-Li Huang, En-Wei Liang}
\email{xiaoli.huang@gznu.edu.cn; lew@gxu.edu.cn}

\author{Ji-Gui Cheng}
\affiliation{Guangxi Key Laboratory for Relativistic Astrophysics, School of Physical Science and Technology, Guangxi University, Nanning 530004, China}

\author{Xiao-Li Huang$^{*}$}
\affiliation{Guizhou Provincial Key Laboratory of Radio Astronomy and Data Processing, School of Physics and Electronic Science, Guizhou Normal Univeristy, Guiyang 550025, China}

\author{Ze-Rui Wang}
\affiliation{College of Physics and Electronic Engineering, Qilu Normal University, Jinan 250200, China}

\author{Jian-Kun Huang}
\affiliation{Guangxi Key Laboratory for Relativistic Astrophysics, School of Physical Science and Technology, Guangxi University, Nanning 530004, China}

\author{En-Wei Liang$^{*}$}
\affiliation{Guangxi Key Laboratory for Relativistic Astrophysics, School of Physical Science and Technology, Guangxi University, Nanning 530004, China}

\begin{abstract}
A hard TeV $\gamma$-ray component excess over the single-zone leptonic model prediction (TeV excess) is observed in the spectral energy distributions (SEDs) of some BL Lacs. Its origin is uncertain. We revisit this issue with four BL Lacs (1ES 0229+200, 1ES 0347--121, 1ES 1101--232, and H2356--309), in which the TeV excess is detected in their intrinsic SEDs. We represent their SEDs with a single-zone leptohadronic model, where radiations of the electrons and protons as well as the cascade electrons produced by the $\gamma\gamma$ and p$\gamma$ interactions within their jets are considered. We show that the observed SEDs below the GeV gamma-ray band are attributed to the synchrotron radiations and self-Compton process of the primary electrons, and the TeV excess is explained with the $\gamma$-ray emission from the p$\gamma$ process via the $\pi^{0}$ decay. The cascade emission of the electrons produced via the $\gamma\gamma$ and p$\gamma$ interactions results in a keV-MeV excess in the SEDs, illustrated as a bump or plateau. This extra photon field enhances the production of TeV photons from the $p\gamma$ process, resulting in a reduction of the proton power by about one order of magnitude. However, the derived powers are still 3--4 orders of magnitude larger than the Eddington limit, being challenged by the current black hole accretion physics. Applying our model to Mrk 421, we propose that synergic observations with current and upcoming TeV and keV-MeV telescopes for its tentative TeV and MeV excesses can give insights to the hadronic process in its jet.
\end{abstract}


\section{Introduction}
\label{sec:intro}

Most TeV $\gamma$-ray emitting active galactic nuclei (AGNs) are BL Lacertaes (BL Lacs)\footnote{TeVCat official website: \url{http://tevcat.uchicago.edu/}}. Their observed broadband spectra energy distributions (SEDs) are characterized as bimodal distribution with peaks at the IR-Optical-X-ray and the MeV-GeV bands \citep{Urry&Padovani1995}. It is generally believed that the SEDs are attributed to leptonic processes of relativistic electrons accelerated within the jets. The peak at the low-energy band is explained as the synchrotron radiations of the electrons, and the high-energy peak results from the Synchrotron-Self-Compton (SSC) scattering process (e.g. \citealt{MaraschiL...1992,Ghisellini&Madau1996}). \cite{ZhangJ...2012} presented a systematical analysis for the SEDs of 24 TeV BL Lacs with the single-zone leptonic model. They showed that most SEDs are indeed well represented with the model.

TeV-selected BL Lacs are the most valuable sources for studying particle accelerations and radiation physics. The obstacle for this purpose is that the extragalactic TeV photons are absorbed by the extragalactic background light (EBL) via the pair production process when traveling through the universe \citep{Gould&Schreder1966}. Interestingly, the intrinsic SEDs of some BL Lacs show an apparent excess in the TeV band (the TeV excess) by correcting the EBL absorption. The excess is an extra hard spectral component that is difficult to be explained with the single-zone leptonic model. For instance, the High Energy Stereoscopic System (HESS) observations show that the intrinsic photon spectrum in an outburst of 1ES 1101-232 during March 2004 to June 2005 is very hard with a power-law index $\sim 1.5$ in the energy range from 0.23 to 4 TeV \citep{HESS_1101_2007}. In the hadronic model, the TeV excess is suggested to be produced through proton synchrotron radiation or the decay of neutral pions produced via the $pp$ or $p\gamma$ interaction process \citep{MannheimK1993}. \cite{Cao&Wang2014} studied the possible hadronic origin of the hard $\gamma$-ray spectrum and suggested that the TeV excess would originate from the decay of neutral pion produced through p$\gamma$ interactions, in which the soft photons are from the electron synchrotron radiation within the jet. \cite{SahuS...2013} proposed that the observed TeV orphan flare of 1ES 1959+650 is also from the photohadronic process. \cite{MastichiadisA...2013} investigated the X-ray and $\gamma$-ray variabilities of Mrk 421 in various leptohadronic scenarios and found that the hadronic model can reproduce the quadratic behavior between X-ray and TeV observations. Alternatively, it has been found that the TeV excess could originate from leptonic processes of different radiation zones. Such as inverse Compton scattering of cosmic microwave background (IC/CMB) photons by electrons in the large scale jet \citep{BottcherM...2008,YanDH...2012}, or SSC process in a more distant region away from the emitting region that is responsible for the outbursts in the keV-GeV band \citep{ZhangJ2009}.


Hadronic models for explaining the gamma-ray emission of AGN were also proposed. These models usually require a jet power being larger than the Eddington luminosity by 1-2 orders of magnitude (e.g. \citealt{BottcherM...2013}). Although a moderate proton power could be enough for modeling the MeV-GeV bump of the SEDs (e.g. \citealt{CerrutiM...2015}), an extremely large jet power, which is larger than the Eddington luminosity by 4-5 orders of magnitude, is required for attributing the TeV excess to the contributions of the $p\gamma$ process (\citealt{Cao&Wang2014}). This sharpens the issue of the breaking Eddington limit. Despite of the Eddington luminosity is not a strict limit, it is a reasonable approximation for the maximum jet power of a blazar (e.g. \citealt{Zdziarski&Bottcher2015}).

The observed TeV excess indicates that the $e^{\pm}$-pair production and its cascade emission within the jet may play a key role in shaping the observed SED in the keV-GeV band \citep{CerrutiM...2015,XueR...2021}. More importantly, the cascade emission should offer an extra photon field for enhancing the very high energy (VHE) gamma-ray emission via the p$\gamma$ process, leading to a reduction of the proton power that may relieve the tension of exceeding the Eddington luminosity to some extent. This letter aims to explore the hadronic origin of the TeV excesses in the framework of the single-zone leptohadronic model by considering radiations of the relativistic electrons and protons together with their cascade process in detail. We present the selected broadband SEDs in Sec.\ref{sec:excess}, and describe our model in Sec.\ref{sec:model}. The results are reported in Sec.\ref{sec:resu}. Conclusion and discussion are given in Sec.\ref{sec:conc}. Throughout the paper, cosmological parameters of $H_{\rm 0}=70{\rm\,km\,s^{-1}\,Mpc^{-1}}$, $\Omega_{\rm m}=0.3$, and $\Omega_{\rm \Lambda}=0.7$ are adopted.

\section{Intrinsic TeV excess in Selected-SEDs of BL Lacs}
\label{sec:excess}

We select some BL Lacs with clear detections of a TeV excess component over the leptonic model prediction in their SEDs. They are taken from \cite{ZhangJ...2012}, in which a large SED sample of 24 TeV BL Lacs is compiled and represented with the single-zone leptonic model. We derive the intrinsic SEDs by correcting the EBL absorption with the model in \cite{FinkeJD...2010}\footnote{The EBL model by \cite{FranceschiniA...2008} is also adopted for double check the EBL corrections in our analysis. The difference of the derived intrinsic spectra induced by the two EBL models is not significant.}. Based on the leptonic model results presented in \cite{ZhangJ...2012}, we finally select the following 4 BL Lacs that have a TeV excess component over the leptonic model prediction with a confidence level of $3\sigma$ at the energy band of $>$0.1 TeV. The $\sigma$ value is estimated as $\sigma=\sum_{\rm i} (|F_{\rm int, i}-F_{\rm t, i}|/\sigma_{\rm i})$, where $F_{\rm int, i}$ and $\sigma_{\rm i}$ are the intrinsic flux and its uncertainty, respectively, and $F_{\rm t, i}$ is the intrinsic flux predicted by the leptonic model. The selected SEDs for the 4 BL Lacs are shown in Fig.\ref{fig:seds} and described as follows.

\begin{itemize}
    \item \textit{1ES 0229+200} (\textbf{$z=0.14$; \citealt{Aharonian...2000}}). This BL Lac is known for its hard spectrum in the TeV band. HESS observations of this source in 2005 and 2006 show a photon spectral index of $\sim 2.5$ in the energy range 500 GeV to $\sim 15$ TeV, yielding a very hard intrinsic spectrum at very high energy (VHE) band \citep{HESS_0229_2007}. Between 2009 October and 2013 January, the VERITAS detects an excess of 489 $\gamma$-ray events from 1ES 0229+200 in the energy range of $0.29-7.6$ TeV, with an average integral flux of $\sim 23.3\times10^{-9}\,{\rm photons\,m^{-2}\,s^{-1}} $\citep{AliuE...2014a}. The long-term observations of the MAGIC, with a total of 265 hr good-quality data from 2010 to 2017, suggested that the intrinsic spectral index of 1ES 0229+200 is $\sim$ 1.8 at 521 GeV \citep{MAGIC_0229_2020}, for which the EBL model of \cite{FranceschiniA...2008} is considered. The hadronic origin of the VERITAS observation has been discussed with various models, for example, the proton-synchrotron and the leptohadronic models \citep{CerrutiM...2015}. In our work, multi-band data include VHE and X-ray observations, which are collected from \cite{HESS_0229_2007} and \cite{ZhangJ...2012}, respectively.
    \item \textit{1ES 0347--121} (\textbf{$z=0.188$; \citealt{Woo...2005}}). It was observed by the HESS between 2006 August and December in the VHE band. The detected photon spectral index is $\sim 3.10$ in the energy range $0.25-3$ TeV, with an integral flux $\sim 2\%$ of that of the Crab Nebula \citep{VHE_0347_2007}. The hadronic origin of this VHE observation has also been studied and discussed \citep{CerrutiM...2015}. The multi-band data used in our intrinsic broadband SED analysis are obtained from \cite{VHE_0347_2007}, which includes the ATOM and SWIFT observations.
    \item \textit{1ES 1101--232} (\textbf{$z=0.186$; \citealt{Wolter...2000}}).  From March 2004 to June 2005, HESS detected 1ES 1101--232 with an excess of 649 photons. This VHE observation shows a very hard intrinsic spectrum with a photon index $\sim 1.5$ from 0.23 to 4 TeV \citep{HESS_1101_2007}. Previous studies have indicated that the VHE emission may originate from the hadronic processes (e.g. \citealt{Cao&Wang2014,CerrutiM...2015}). In this work, the multi-band data of the broadband SED including the HESS observation, as well as the semi-simultaneous X-ray observation of the XMM-Newton and RXTE are taken from \cite{Xray_1101_2007}.
    \item \textit{H2356--309} (\textbf{$z= 0.165$; \citealt{Bersanelli...1992}}). The VHE observations of H2356--309 have been reported twice by the HESS. The first one was from 2004 June to December with a photon index $\sim 3.09$ in the energy range from 0.2 to 1.3 TeV \citep{HESS_2356_2006}. The second one was observed over a long-term period, from 2004 to 2007, yielding the integral flux $\sim 3.06\times10^{-12}\,{\rm photons\,cm^{-2}\,s^{-1}}$ above 240 GeV and the photon index $\sim 3.06$ in the energy range from 200 GeV to 2 TeV \citep{HESS_2356_2010}. It has been found that the H2356--309 location of the former is coincident with the error circles of three IceCube events, and the VHE emission may be from the hadronic processes \citep{SahuS&MirandaLS_2015}. The multi-band data used in this paper are obtained from \cite{HESS_2356_2010}, incorporating the simultaneous optical/UV and X-ray observations from the XMM-Newton, NRT radio observation, and ATOM optical observation.
\end{itemize}

\section{Model}
\label{sec:model}

We employ a single-zone leptohadronic model to explain the broadband SEDs of the selected BL Lacs. The bimodal SED from the radio to the MeV-GeV $\gamma$-ray band are attributed to the synchrotron radiations of electrons accelerated in the jet. The TeV excess is explained as the production of the interaction between accelerated protons and photons (p$\gamma$ process). Cascade emission is considered following that presented by \cite{BottcherM...2013}. The injected cascade electrons include electrons produced via the internal $\gamma\gamma$ absorption for photon fields of both the leptonic and hadronic processes, and electrons from charged pion decay in the $p\gamma$ process (e.g. \citealt{Atoyan&Dermer2003,Romero&Vila2008}), as well as electrons resulting from the Bethe-Heitler process (e.g. \citealt{ChodorowskiMJ...1992,Kelner&Aharonian2008}). We describe the model in the following.

\subsection{The Leptonic Radiations}
\label{subsec:lept}

 Our leptonic model is adopted as \cite{ZhangJ...2012}. The emitting region is assumed to be a single-zone sphere with a radius $R_{\rm b}$ moving with a bulk Lorentz factor $\Gamma$ (or a velocity in unit of the light speed, $\beta$). $R_{\rm b}$ can be estimated with the minimum variability timescale, $R_{\rm b} = c \delta_{\rm D} \Delta t / (1+z)$, where $\delta_D= 1/\Gamma(1-\beta \cos\theta)$ is the Doppler factor with an viewing angle of $\theta$ to the jet axis. For BL Lacs, the jet orientation is close to the line of sight, and thus $\delta_{\rm D} \sim \Gamma$. The primary electron spectrum is taken as a broken power-law function with indices $p_1$ and $p_2$ breaking at $\gamma_{\rm e, b}$ in the range of [$\gamma_{\rm e,min}$, $\gamma_{\rm e,max}$].

The bimodal SED feature in the radio-optical-X-ray-GeV $\gamma$-ray band is attributed to the synchrotron radiation and SSC process of the relativistic electrons. Our calculation of the synchrotron process is based on an approximation method \citep{naima_syn_2010}, for which the synchrotron self-absorption \citep{Mkn501_2001} is considered. The SSC process is calculated with analytical approximations \citep{naima_IC_2014}. The internal $\gamma\gamma$ absorption \citep{Finke&Dermer2008} is yielded, where the high-energy and low-energy photons are from the SSC and synchrotron radiations, respectively. The radiative cooling of electrons can be estimated by the energy density of the magnetic field and the synchrotron photon field, in which the Klein-Nishina effect is incorporated \citep{KN_effect_2005}.

\subsection{The Hadronic Radiation}
\label{subsec:hadr}

 We consider only the radiative cooling of the protons via the $p\gamma$ process. There are two main channels of this process: (1) The Bethe--Heitler (B--H) process, ${\rm p} + \gamma \rightarrow {\rm p} + {\rm e^{-}} + {\rm e^{+}}$ for photons with a threshold energy of $\sim 1$ MeV in the proton rest frame \citep{Kelner&Aharonian2008}; (2) The photomeson process of single-pion pass ${\rm p} + \gamma \rightarrow {\rm p} + {\rm a}\pi^{0}+ {\rm b}(\pi^{+}+\pi^{-})$ and multi-pion pass ${\rm p} + \gamma \rightarrow {\rm n} + \pi^{+} + {\rm a}\pi^{0} + {\rm b}(\pi^{+}+\pi^{-})$ \citep{Romero&Vila2008} for photons with a  threshold energy of $\sim 145$ MeV in the proton rest frame \citep{Atoyan&Dermer2003}. A $\pi^{0}$ particle decays to two energetic photons, which may result in the observed TeV excess. The electrons generated through the two channels may be cooled via the synchrotron radiation and SSC process. This electromagnetic cascade emission may modify the SED in the radio-optical-X-ray-GeV band.

Our calculations of the $p\gamma$ process are based on the semi-analytical method under the framework of \cite{Atoyan&Dermer2003} and \cite{Romero&Vila2008}. The collision rate of a proton with the Lorentz factor $\gamma_{\rm p}$ through the $p\gamma$ process with a soft photon field can be expressed as

\begin{equation}
    \omega_{\rm p\gamma,i}(\gamma_{\rm p}) = \frac{c}{2\gamma_{\rm p}^{2}} \int_{\frac{\epsilon_{\rm th,i}^{'}}{2\gamma_{p}}}^{\infty} d\epsilon \frac{n_{\rm ph}(\epsilon)}{\epsilon^{2}} \int_{\epsilon_{\rm th,i}^{'}}^{2\epsilon\gamma_{\rm p}} d\epsilon^{'} \sigma_{\rm p\gamma,i}(\epsilon^{'}) \epsilon^{'},
\end{equation}
where the subscript ${\rm i} = {\rm e}, \pi$ represents for the B--H and photomeson processes respectively, $n_{\rm ph} (\epsilon)$ is the number density of photons with energy $\epsilon$ (the synchrotron radiation of electrons in our case), $\epsilon^{'}$ and $\epsilon_{\rm th,i}^{'}$ are the photon energy and photon threshold energy in the proton rest frame, and $\sigma_{\rm p\gamma,i}$ is the cross section. Thus, the cooling rate can be given by
\begin{equation}
\label{eq:pcooling}
    \tau_{\rm p\gamma,i}^{-1}(\gamma_{\rm p}) = \frac{c}{2\gamma_{\rm p}^{2}} \int_{\frac{\epsilon_{\rm th,i}^{'}}{2\gamma_{p}}}^{\infty} d\epsilon \frac{n_{\rm ph}(\epsilon)}{\epsilon^{2}} \int_{\epsilon_{\rm th,i}^{'}}^{2\epsilon\gamma_{\rm p}} d\epsilon^{'} \sigma_{\rm p\gamma, i}(\epsilon^{'})K_{\rm p\gamma, i}(\epsilon^{'})\epsilon^{'},
\end{equation}
where $K_{\rm p\gamma, e}$ is the inelasticity factor. The cross section $\sigma_{\rm p\gamma,e}$ and inelasticity $K_{\rm p\gamma, e}$ can be found in \cite{ChodorowskiMJ...1992}. For the photomeson process, the cross section and inelasticity can be approximated as
\begin{equation}
\sigma_{\rm p\gamma,\pi}(\epsilon^{'}) \approx \left\{
    \begin{array}{lll}
    3.4\times10^{-28}\,{\rm cm^{-2}}, && 200\,{\rm MeV} \leqslant \epsilon^{'} \leqslant 500\,{\rm MeV} \\
    1.2\times10^{-28}\,{
    \rm cm^{-2}}, && \epsilon^{'} > 500\,{\rm MeV},
    \end{array}\right.
\end{equation}
\begin{equation}
K_{\rm p\gamma,\pi}(\epsilon^{'}) \approx \left\{
    \begin{array}{lll}
    0.2, && 200\,{\rm MeV} \leqslant \epsilon^{'} \leqslant 500\,{\rm MeV} \\
    0.6, && \epsilon^{'} > 500\,{\rm MeV},
    \end{array}\right.
\end{equation}
in which the two energy ranges are correspond to single-pion and multi-pion channels, respectively.

The proton distribution in the energy range [$E_{\rm p,min}$, $E_{\rm p,max}$] is assumed as an exponential cutoff power-law with an index $\alpha$ and an cutoff energy $E_{\rm p,cut}$, i.e.,
\begin{equation}
    N_{\rm p}(E_{\rm p}) = N_{0} \left(\frac{E_{\rm p}}{1\,{\rm eV}}\right)^{-\alpha} {\rm exp}\left(-\frac{E_{\rm p}}{E_{\rm p,cut}}\right),\qquad E_{\rm p,min} \leqslant E_{\rm p} \leqslant E_{\rm p,max}.
\end{equation}

The electron energy produced in the B--H process depends on the proton energy as $E_{\rm e} = \frac{1}{2} \overline{K}_{\rm p\gamma,e}E_{\rm p}$, and its productivity is
 \begin{equation}
 \label{eq:BH}
    Q_{\rm BH, e}(E_{\rm e}) = \frac{4}{\overline{K}_{\rm p\gamma,e}} N_{\rm p}(\frac{2}{\overline{K}_{\rm p\gamma,e}}E_{\rm e}) \omega_{\rm p\gamma, e}(\frac{2}{\overline{K}_{\rm p\gamma,e}}\frac{E_{\rm e}}{m_{\rm p}c^{2}}).
\end{equation}
The mean inelasticity $\overline{K}_{\rm p\gamma,e}$ can be approximated to $2m_{\rm e}/ m_{\rm p}$ \citep{Romero&Vila2008}. Due to that electrons produced through a single proton-photon interaction in the B--H process has a broad energy distribution, the productivity calculated through the mean inelasticity and Eq.(\ref{eq:BH}) is not accurate enough. Hence, we adopt an analytic method introduced by \cite{Kelner&Aharonian2008} to calculate the electron production. This method is based on the differential cross section of the process and can provide better accuracy.

Mesons ($\pi^{0}$ and $\pi^{\pm}$) produced in the p$\gamma$ interaction would inherit about 20\% of the energy from the collided proton and further pass it to photons and electrons, $E_{\gamma} \approx 0.1 E_{\rm p}$ and $E_{\rm e} \approx 0.05 E_{\rm p}$. The photon emissivity from $\pi^{0}$ decay and the electron emissivity from $\pi^{\pm}$ decay can be calculated as
\begin{equation}
    Q_{p\gamma, \gamma}(E_{\gamma}) = 20 N_{\rm p}(10E_{\gamma}) \omega_{\rm p\gamma,\pi}(10E_{\gamma}) n_{\pi^{0}}(10E_{\gamma}),
\end{equation}
\begin{equation}
    Q_{\rm p\gamma, e}(E_{\rm e}) = 20 N_{\rm p}(20E_{\rm e}) \omega_{\rm p\gamma,\pi}(20E_{\rm e}) n_{\pi^{\pm}}(20E_{\rm e}),
\end{equation}
where $n_{\pi^{0}} \approx 0.5p_{1} + p_{2}$ and $n_{\pi^{\pm}} \approx 0.5p_{1} + 2p_{2}$ are defined as the mean number of $\pi^{0}$ and $\pi^{\pm}$ created per collision, respectively. $p_{1}$ is the possibility of the p$\gamma$ interaction through the single-pion pass and $p_{2} = 1 - p_{1}$ is for the multi-pion pass. They can be derived by defining the mean inelasticity as
\begin{equation}
    \overline{K}_{\rm p\gamma,\pi} = \tau_{\rm p\gamma,\pi}^{-1} \omega_{\rm p\gamma,\pi}^{-1} = 0.2p_{1} + 0.6(1-p_{1}).
\end{equation}

The specific luminosity of the gamma-ray photons from the $\pi^{0}$ decay in the comoving frame is given by
\begin{equation}
  L_{\rm E}(E_\gamma)=4\pi E_\gamma Q_{\rm p\gamma, \gamma}(E_{\gamma}) V \times \frac{1-e^{-\tau_{\gamma\gamma}}}{\tau_{\gamma\gamma}},
\end{equation}
where $V = 4 \pi R_{\rm b}^{3} / 3$ is the volume of the radiating region and $\tau_{\gamma\gamma}$ is the optical depth of the internal $\gamma\gamma$ interaction \citep{Finke&Dermer2008}. Without considering the EBL absorption, the intrinsic specific flux in the observed frame is $F_{\rm E}(E_{\rm \gamma, obs}) = (1+z) \delta_{\rm D}^{3} L_{\rm E}(E_\gamma)/4\pi D_{\rm L}^{2}$, where $E_{\rm \gamma, obs}=E_\gamma\delta_{\rm D}/(1+z)$, $D_{\rm L}$ is the luminosity distance.

%

\subsection{The Cascade Process}
Generally, the pair cascade process is initiated by the internal $\gamma\gamma$ interaction of existing photon fields inside the radiation region, creating a loop between electron generation, cooling, and escaping. The system eventually reaches a temporary equilibrium and returns a stationary secondary electron distribution. The cascade process is calculated in a time-independent method \citep{BottcherM...2013}, and the Fokker--Planck equation is written as
\begin{equation}
\label{eq:cas}
    \frac{\partial}{\partial \gamma_{\rm e}}(\dot{\gamma}_{\rm e} N_{\rm e}^{\rm cas}) = Q_{\rm e}(\gamma_{\rm e}) + \dot{N}_{\rm e}^{\gamma\gamma}(\gamma_{\rm e}) + \dot{N}_{\rm e}^{\rm esc}(\gamma_{\rm e}),
\end{equation}
where $Q_{\rm e}$ is the sum of the B--H and photomeson electrons in the p$\gamma$ interaction, $\dot{N}_{\rm e}^{\gamma\gamma}$ is the electron injection from the internal $\gamma\gamma$ interaction, $\dot{N}_{\rm e}^{\rm esc}$ is the escape term, and $N_{\rm e}^{\rm cas}$ is the final electron distribution. The particle escape can be estimated as $\dot{N}_{\rm e}^{\rm esc}(\gamma_{\rm e}) = N_{\rm e}^{\rm cas}(\gamma_{\rm e}) / t_{\rm esc}$, where \textbf{$t_{\rm esc} = R_{\rm b} / c$}. $\dot{N}_{\rm e}^{\gamma\gamma}(\gamma_{\rm e})$ is given by
\begin{equation}
    \dot{N}_{\rm e}^{\gamma\gamma}(\gamma_{\rm e}) = f_{\rm abs}(\epsilon_{1})(\dot{N}_{\epsilon_{1}}^{\rm 0} + \dot{N}_{\epsilon_{1}}^{\rm syn} + \dot{N}_{\epsilon_{1}}^{\rm SSC}) + f_{\rm abs}(\epsilon_{2})(\dot{N}_{\epsilon_{2}}^{\rm 0} + \dot{N}_{\epsilon_{2}}^{\rm syn} + \dot{N}_{\epsilon_{2}}^{\rm SSC}),
\end{equation}
where $\dot{N}^{\rm 0}$ is the initial $\gamma$-rays from the $\pi^{0}$ decay, $\dot{N}^{\rm syn}$ and $\dot{N}^{\rm SSC}$ are the synchrotron and SSC radiations of $N_{\rm e}^{\rm cas}$, $\epsilon_{1}$ and $\epsilon_{2}$ are the energies of photons to produce $\gamma_{\rm e}$ electrons via the pair production, and $f_{\rm abs}$ is the internal absorption coefficient. $\epsilon_{1}$ and $\epsilon_{2}$ are estimated as $\epsilon_{1}=\gamma_{\rm e} / f_{\gamma}$ and $\epsilon_{2} = \gamma_{\rm e} / ( 1 - f_{\gamma})$, where $f_\gamma$ is a energy fraction of the two photons, which is taken as 0.9 following \cite{BottcherM...2013}.

\section{Results}
\label{sec:resu}
\begin{figure}[htbp]
    \centering
    \includegraphics[width=0.45\textwidth, angle=0]{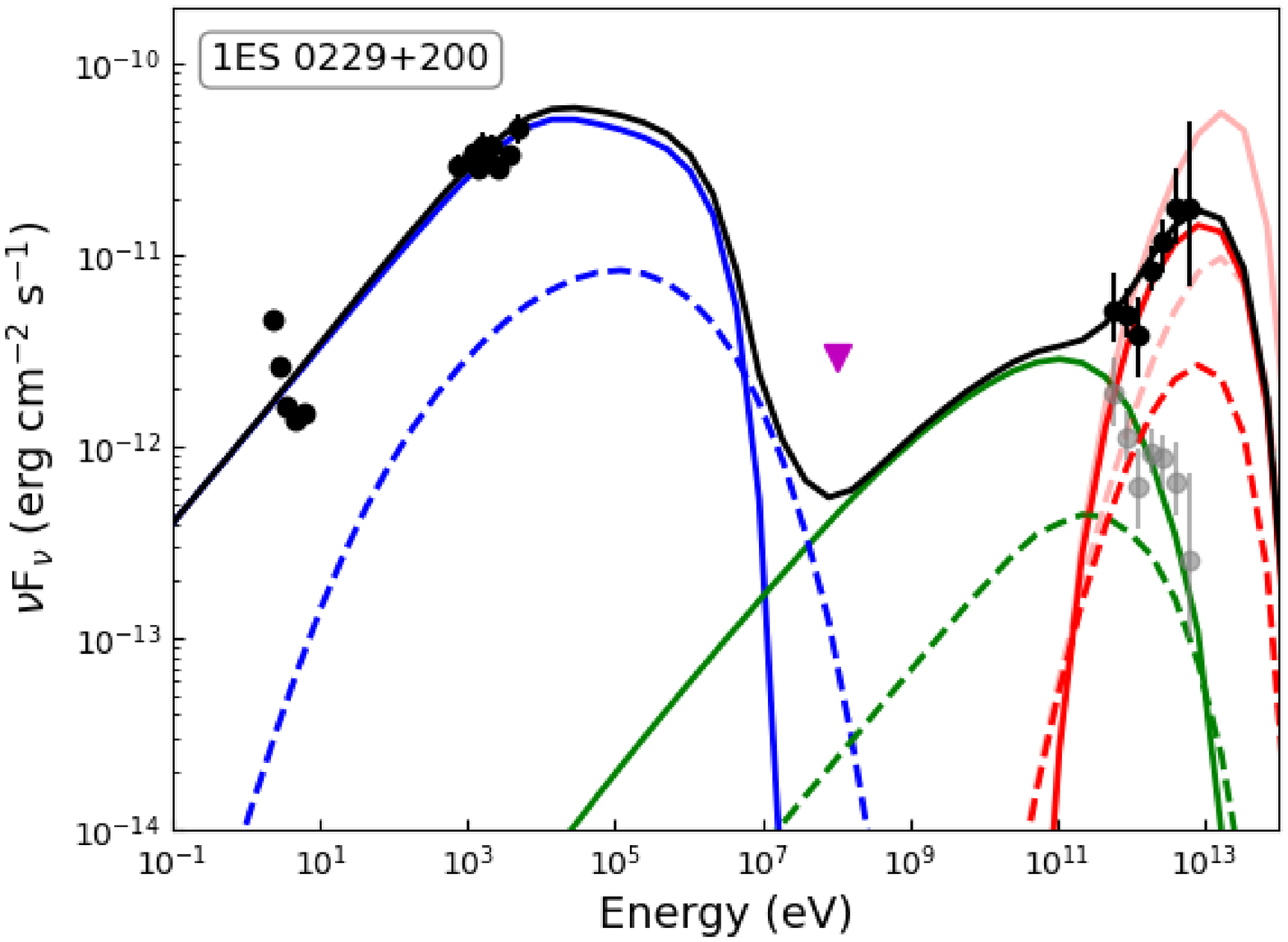}
    \includegraphics[width=0.45\textwidth, angle=0]{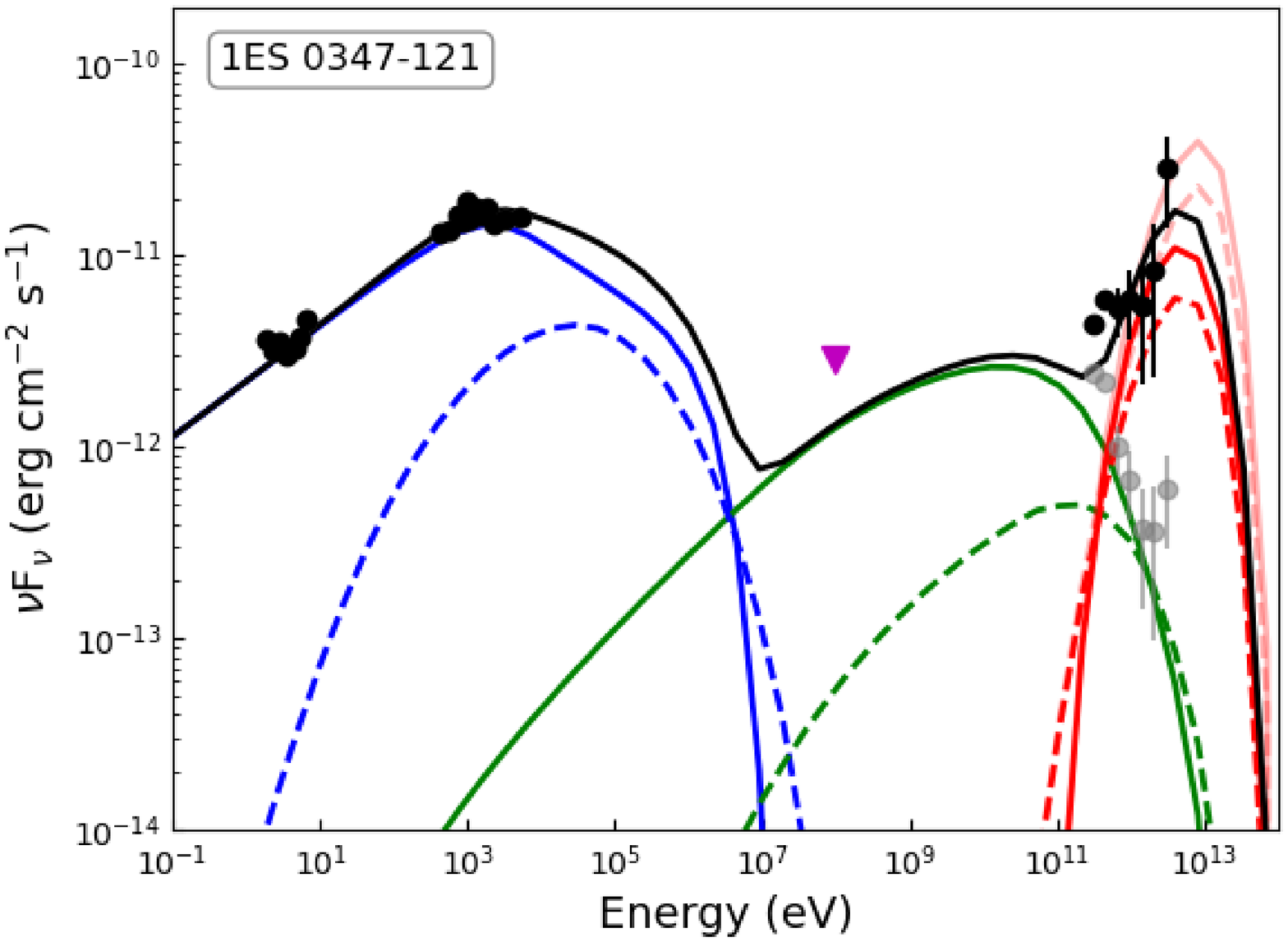}
    \includegraphics[width=0.45\textwidth, angle=0]{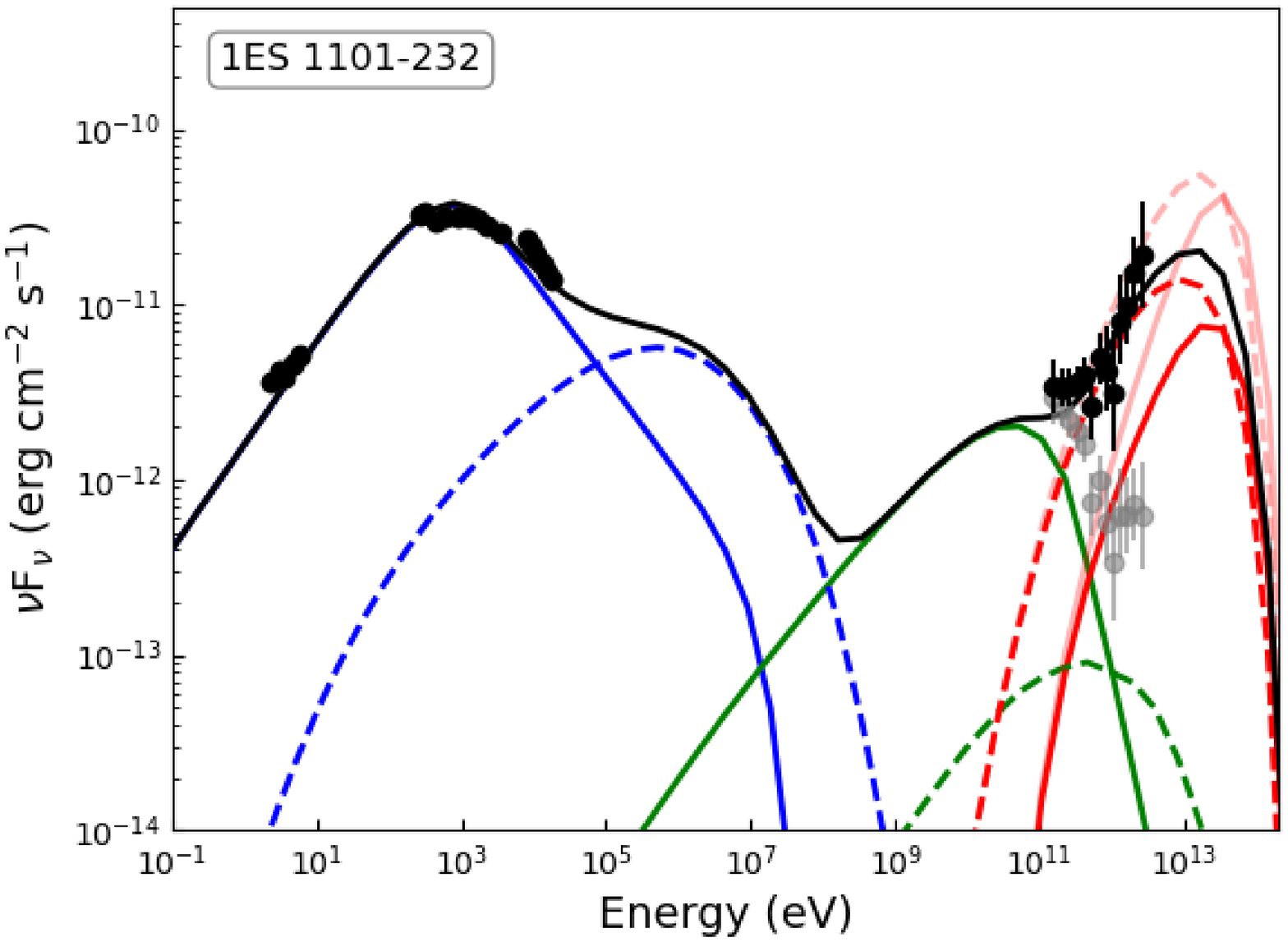}
    \includegraphics[width=0.45\textwidth, angle=0]{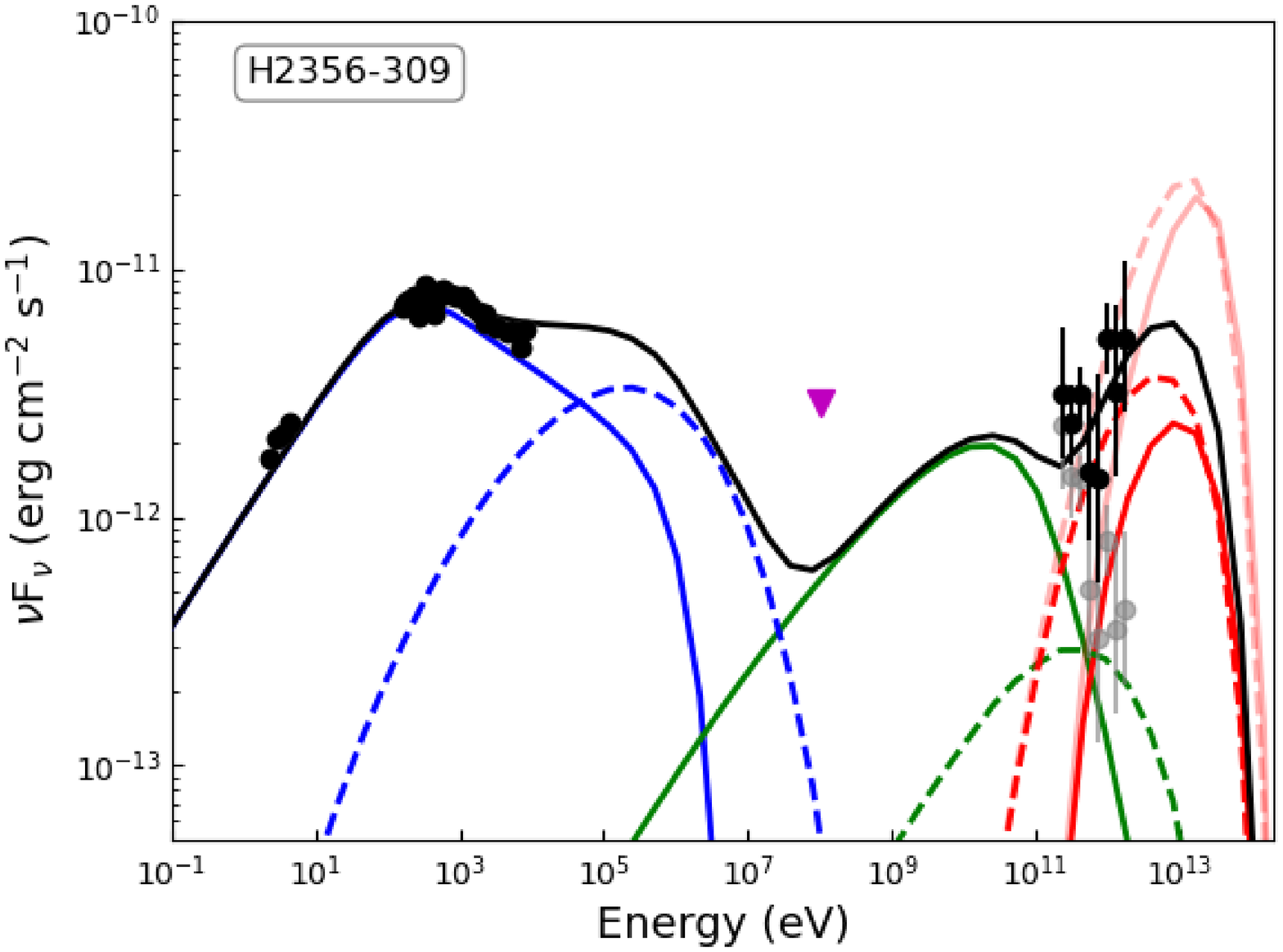}
    \caption{The EBL-absorbtion-corrected SEDs (solid black dots with error bars) and our leptohadronic model results (solid black curves). The gray points are the observed data without the EBL absorption correction. The purple inverted triangles indicate the upper limit of the \textit{Fermi/LAT} observations from \cite{TavecchioF...2010}. Color lines represent the different radiation processes: blue for synchrotron, green for SSC, and red for $\pi^{0}$-decay. Solid lines are for the emission of the primary electrons and dashed lines are for the emission of the cascade electrons. The hadronic emission without considering the internal $\gamma\gamma$ absorption are also drawn with red translucent solid/dashed lines for demonstration.}
\label{fig:seds}
\end{figure}

Assuming that the protons are accelerated by the relativistic shocks via the Fermi acceleration mechanism, the energy spectral index of proton distribution is fixed at $p_{\rm p}=2.2$ in our calculations. In the leptohadronic model, contributions from the following spectral components shape the SEDs, i.e., (1) the primary leptonic emission components, including the synchrotron and SSC emission components from the primary electron population, (2) the primary VHE $\gamma$-ray component from the p$\gamma$ process via the $\pi^{0}$ decay, in which photons are from the primary synchrotron radiations, (3) the cascade leptonic component, including the synchrotron and SSC emission components from the cascade electrons produced via the internal $\gamma\gamma$ interaction and the $p\gamma$ interaction, and (4) the cascade VHE $\gamma$-ray component from the cascade p$\gamma$ process, in which photons are from the cascade synchrotron radiations. Our numerical results are shown in Fig.\ref{fig:seds} and the model parameters are reported in Tab.\ref{tab:paras}. One can observe that the model can well represent the data. The parameters of the leptonic model part are generally consistent with those reported in the literature (e.g. \citealt{ZhangJ...2012,TavecchioF...2010,VHE_0347_2007,Xray_1101_2007,HESS_2356_2010}). The cut-off energy ($E_{\rm p, cut}$) of protons is typically $10$ TeV.

As shown in Fig.\ref{fig:seds}, the synchrotron radiations of the primary electron population overwhelmingly dominate the SEDs in the optical-soft-X-ray band ($1 \sim 10^4$ eV), and the SSC components of the primary electron population peak at around 10 GeV for the four BL Lacs. The cascade synchrotron radiation bump peaks at around 0.1 MeV. The peak fluxes of both the synchrotron radiation and the SSC emission components of the cascade electron population are lower than that of the primary electron population by a factor of $2\sim 10$. The combinations of the primary and cascade synchrotron emission components shape the SEDs as a distinct hump or a plateau in the keV-MeV energy band, as shown up in the SEDs of 1ES 0347--121, 1ES 1101--232, and H2356--309. Such a keV-MeV excess would be a distinct feature for the hadronic process.

The TeV $\gamma$-rays are contributed by the $\pi^0$-decay process in the primary (or cascade) p$\gamma$ interactions, in which the soft $\gamma$-ray photon fields are the synchrotron radiations of the primary (or cascade) electron population. Both the primary and cascade SSC emission decrease rapidly at high energies because of the Klein--Nishina effect, making negligible contributions to the TeV excess. The $\pi^0$-decay process completely dominates the TeV $\gamma$-ray flux, although the internal $e^\pm$-pair production is very significant for these VHE $\gamma$-ray photons. Since we have fixed the energy spectral index of the injected proton distribution, the shape of the hadronic spectrum depends largely on the soft photon fields. Note that the photon threshold energy of the p$\gamma$ interaction is $\epsilon_{\rm th} > 145\,{\rm MeV} / 2\gamma_{\rm p}$. Thus, the soft photons should be at least as energetic as $10$ keV for generating the TeV $\gamma$-rays. For 1ES 0229+200 and 1ES 0347--121, the peak flux of the TeV $\gamma$-rays from the primary p$\gamma$ interactions is larger than that from the cascade p$\gamma$ interactions. Inversely, the TeV $\gamma$-ray peak flux in 1ES 1101--232 and H2356--309 is mainly contributed by the cascade p$\gamma$ interactions. The intrinsic total peak flux of the TeV excess is $F_{\rm int, p}\sim 1\times10^{-11}\,{\rm erg\,cm^{-2}\,s^{-1}}$, comparable to the synchrotron peak flux of the primary electrons.

\section{Discussion}

\subsection{The Issue of Breaking Eddington Limit}

\begin{figure}[htbp]
    \centering
    \includegraphics[width=0.35\textwidth, angle=0]{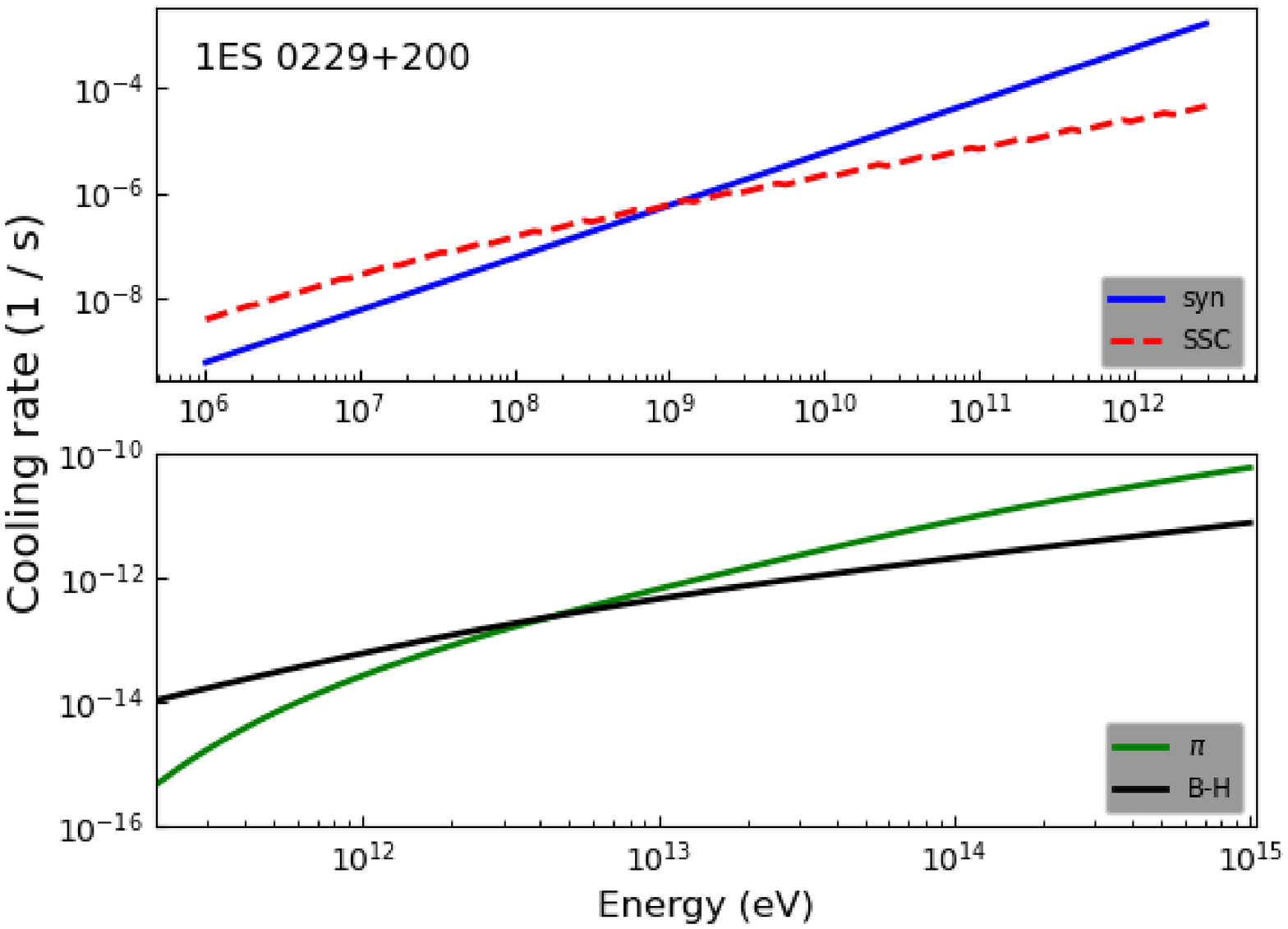}
    \includegraphics[width=0.35\textwidth, angle=0]{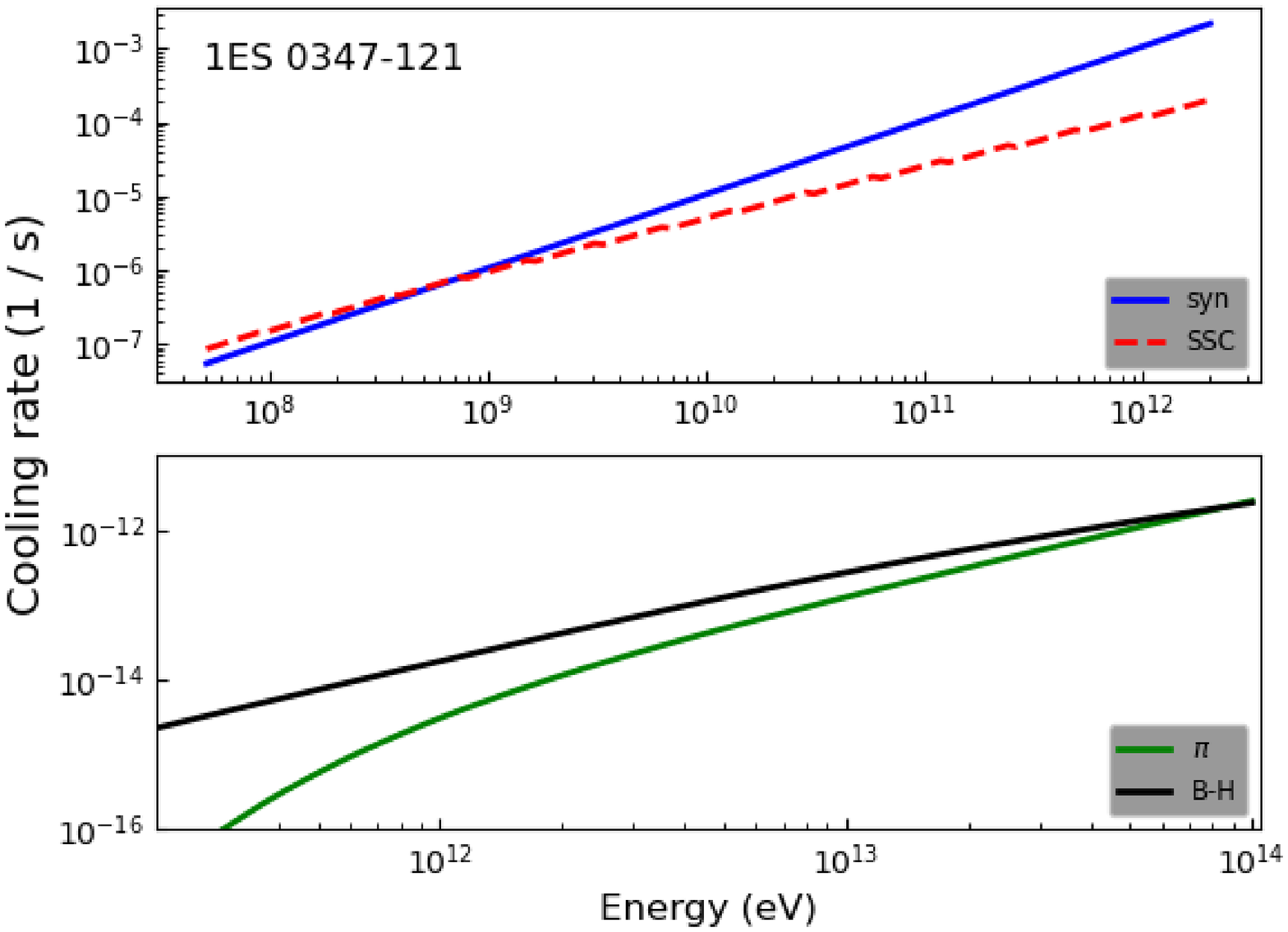}
    \includegraphics[width=0.35\textwidth, angle=0]{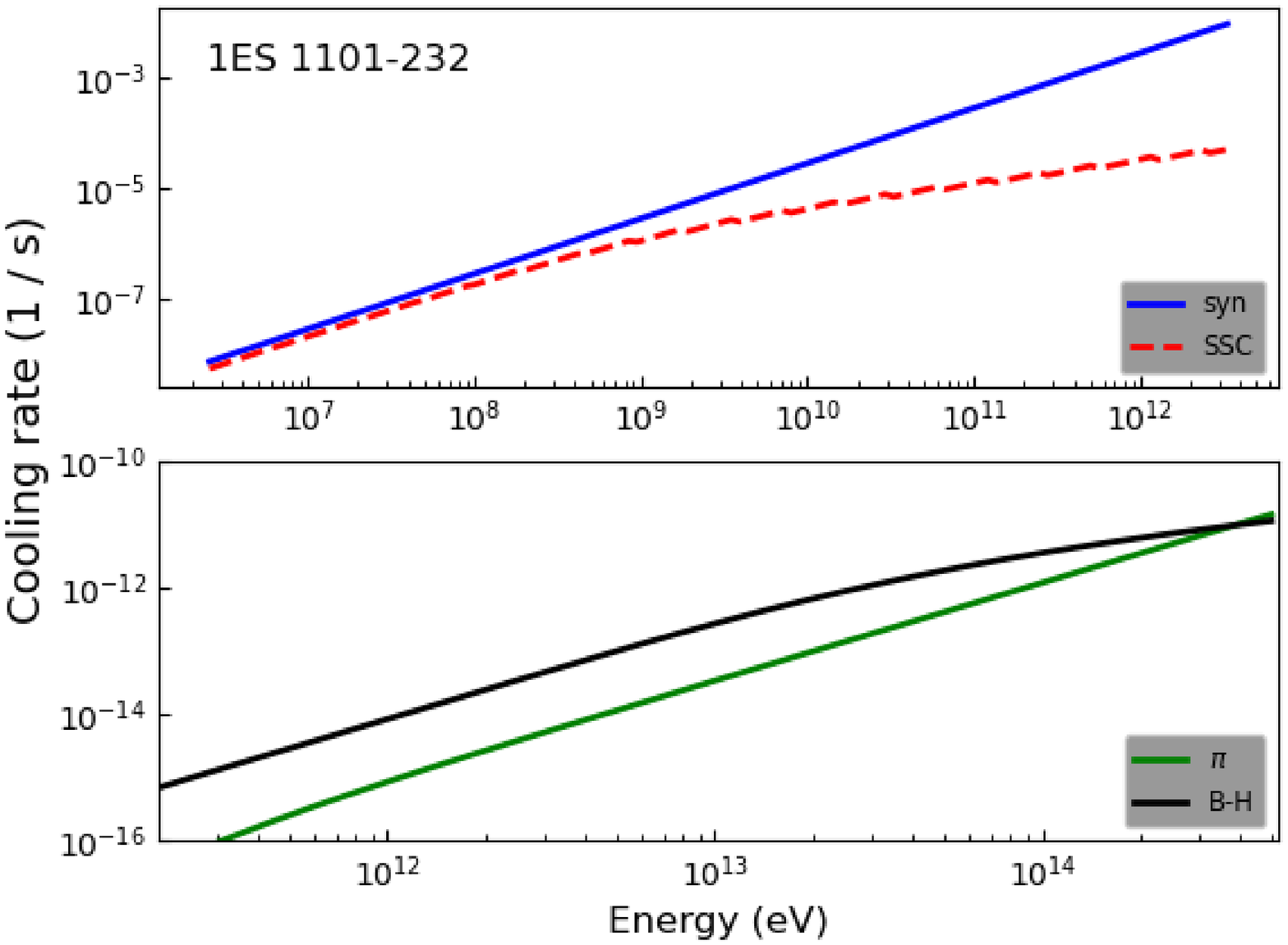}
    \includegraphics[width=0.35\textwidth, angle=0]{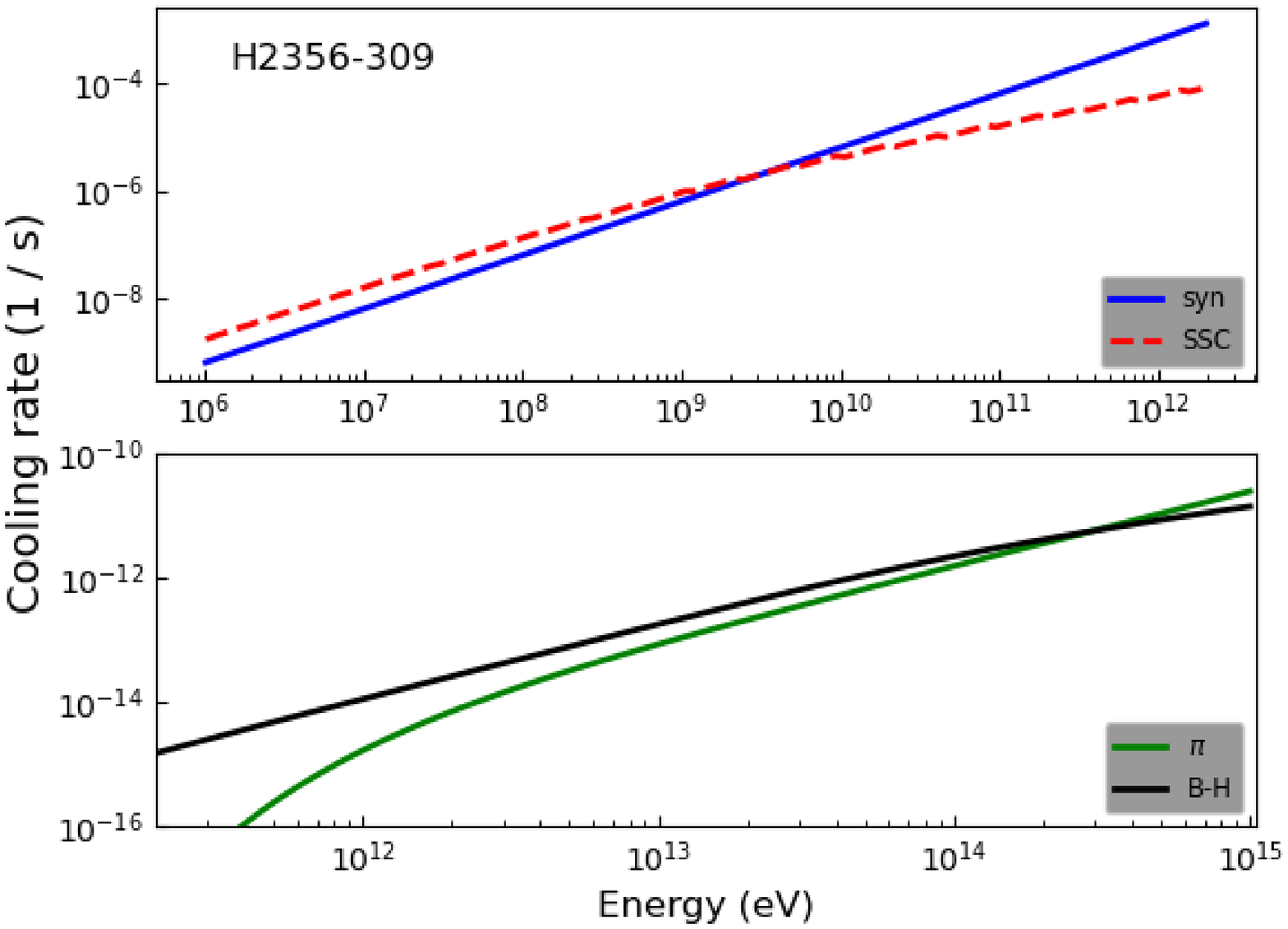}
    \caption{The cooling rates of the primary synchrotron and SSC as well as the primary p$\gamma$ components. The upper panel and the lower panel are the cooling rates of the primary electrons and protons for the 4 BL Lacs.}
\label{fig:coolings}
\end{figure}

The derived jet power breaks the Eddington luminosity limit is a long-standing issue in single-zone hadronic (leptohadronic) models (e.g. \citealt{BottcherM...2013}). This issue is sharpened while explaining the TeV excess with gamma-rays generated through the p$\gamma$ process \citep{Cao&Wang2014} due to the inefficient radiative cooling of the protons. We compare the cooling rates of protons and electrons in our model in Fig.\ref{fig:coolings}. It is found that the primary electrons with energy $<10^{9} \sim 10^{10}$ eV are efficiently cooled via the SSC process, and the cooling of electrons with energy $\gtrsim 10^{10}$ eV is dominated by the synchrotron radiations due to that the Klein-Nishina effect reduces the cross section of high energy electrons. The cooling rates of the protons in the energy range from 1 TeV to 100 TeV via the B--H and photomeson processes are about $10^{-15}\sim 10^{-12}\,{\rm s^{-1}}$. In most cases, the photomeson process is less efficient than the B--H process, which has a much lower threshold photon energy. However, the B--H process has a limited influence in calculating the cascade process for relatively small cross-sections and inelasticity. Electron injections from the internal pair production and the charged pion decay are the major parts that decide the stationary electron distribution of the cascade process.

It was also proposed that the TeV excess is resulted from the $p\gamma$ process with an extremely dense photon field \citep{SahuS...2013}. However, how to form such a photon field is a great challenge in this model. In our analysis, the cascade synchrotron photon field is considered in the p$\gamma$ process to enhance the production of VHE photons. We estimate the proton powers of the sources as $P_{\rm p}=\pi R_{\rm b}^{2} \Gamma^{2} c U_{\rm p}$ with the parameters reported in Tab.\ref{tab:paras}. Our results are $P_{\rm p}=2.19\times 10^{50}$, $4.31\times 10^{51}$, $3.06\times 10^{51}$, and $9.85\times 10^{50}$ ${\rm erg\,s^{-1}}$ for 1ES 0229+200, 1ES 0347--121, 1ES 1101--232, and H2356--309, and their Eddington luminosities are  $2.19\times 10^{47}$, $5.63\times 10^{46}$, $1.26\times 10^{47}$, and $5.02\times 10^{46}$ ${\rm erg\,s^{-1}}$, respectively\footnote{Since no estimated central black hole mass of 1ES 1101--232 is available, we use an average black hold mass $10^{9} M_{\odot}$ instead.}. Note that the derived proton powers in our model are lower than that without considering the cascade synchrotron photon field by one order of magnitude, i.e., $3.06\times 10^{51}$ vs. $3.10\times 10^{52}$ ${\rm erg\,s^{-1}}$ of 1ES 1101--232 for instance. However, they are still 3-4 orders of magnitude larger than the Eddington luminosities. 

\subsection{Magnetic Field Strength and the Proton Acceleration}

In our leptohadronic model, the magnetic field strength is about $0.1 \sim 1$ G, comparable to that used in leptonic models for explaining the SEDs below the TeV band, but 2-3 orders of magnitude smaller than that required in the proton-synchrotron hadronic model for representing the MeV-GeV gamma-ray bump (e.g. \citealt{BottcherM...2013,CerrutiM...2015}). We examine whether the proton can be sufficiently accelerated under the relatively small magnetic field strength. Assuming $B=0.5$ G, the Larmor radius of a 100 TeV proton is $\sim 6.67\times10^{11}\,{\rm cm}$, which is still smaller than the typical emitting region size by 5-6 orders of magnitude. Therefore, the magnetic field strength used in our model should be sufficient for accelerating the proton to relativistic in the radiating region.

The particle acceleration mechanism in BL Lacs is still uncertain. In our analysis, we assume that the charged particles are accelerated by the Fermi acceleration mechanism via the relativistic shocks, and the proton distribution $N_{\rm p}(E_{\rm p}) \propto E_{\rm p}^{-2.2}$ is adopted. Note that particles accelerated through magnetic reconnection may result in a harder particle distribution (e.g. \citealt{ZhuYK...2016} and reference therein). The number density of the low energy end of the proton distribution can be significantly reduced, leading to a decrease of the required proton power in our model. We test this scenario with 1ES 1101--232 by taking the proton distribution index as -1.5. The SED is still can be well represented, and the derived $P_p$ is reduced by a factor of 4.5, changing from $3.06\times 10^{51}$ to $6.96\times 10^{50}$ ${\rm erg\,s^{-1}}$. Meanwhile, narrowing the energy range, i.e., setting a larger $E_{\rm p, min}$, for the proton distribution can also help to reduce the power. The $E_{\rm p, min}$ value is fixed at 10 GeV in our calculations for the 4 BL Lacs. By setting $E_{\rm p, min} = 100\,{\rm GeV}$, the derived proton power changes from $3.06\times 10^{51}$ to $1.58\times 10^{51}$ ${\rm erg\,s^{-1}}$ for 1ES 1101--232. Unfortunately, neither of the approaches we have tested can lower the jet power to an acceptable level concerning the Eddington luminosity that the supper-Eddington issue still exists.

\subsection{Synergy of the TeV excess and keV-MeV excess as a probe for the hadronic process}

Our analysis shows that the flux of the cascade synchrotron emission peaks at the keV-MeV energy band. Since it is usually lower than the primary synchrotron flux in the keV band, it may be featured as a keV-MeV excess (a bump or plateau) in the observed SEDs. The TeV excess and the corresponding keV-MeV excess would be promising probes for the hadronic process. Nearby Mrk 421 ($z=0.031$; \citealt{Punch...1992}) is the best candidate to verify this speculation. So far, Mrk 421 has been detected in the TeV band by the H.E.S.S., VERITAS, ARGO-YBJ, and MAGIC telescopes. It was detected during 2003-2004 with an outburst of a peak flux $\sim 135$ mCrab in the X-ray band and $\sim 3$ Crab in the $\gamma$-ray band \citep{Mkn421_2005}. A tentative TeV excess with large error bars is presented in its SED, as shown in Fig.\ref{fig:Mrk421}. The possible hadronic explanation of the VHE observations has already been discussed (e.g. \citealt{MastichiadisA...2013,ZechA...2017}). Here we represent the SED with our leptohadronic model, and the result is also shown in Fig.\ref{fig:Mrk421}. Same as the other four BL Lacs, the TeV excess of Mrk 421 is mainly contributed by $\pi^{0}$ decay. Interestingly, \cite{Nandikotkur...2008} reported an intriguing convex break at 235 MeV in the SED of Mrk 421 observed with CGRO/EGRET. It is possible that the keV-MeV excess moves in the EGRET energy band during the EGRET observation campaign, shaping the convex break.

\begin{figure}[htbp]
    \centering
    \includegraphics[width=0.45\textwidth, angle=0]{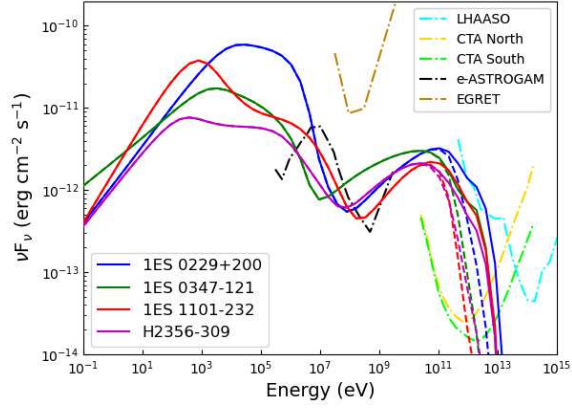}
    \caption{Examination of the detection capability of some instruments for the TeV and MeV excesses in the TeV and MeV bands. The solid lines are our model results presented in Fig.\ref{fig:seds} by considering the EBL-absorption. The dashed lines mark the EBL-absorbed SSC component of the primary electron population. The one-year sensitivities of the instruments are plotted with dash-dotted lines in different colors as marked in the plot.}
    \label{fig:instru&seds}
\end{figure}

\begin{figure}[htbp]
    \centering
    \includegraphics[width=0.45\textwidth, angle=0]{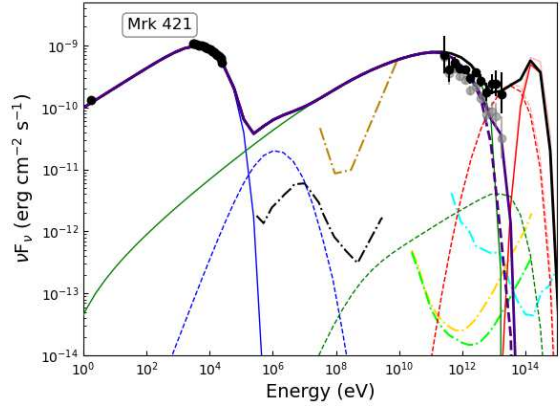}
    \caption{Intrinsic (black dots) and observed (grey dots) SEDs of Mrk 421 and our theoretical modeling (showing with the same symbols as that in Fig.\ref{fig:seds}) together with examination of detection capability with some instruments for the TeV and MeV excesses in the TeV and MeV bands (showing with the same symbols as that in Fig.\ref{fig:instru&seds}).}
    \label{fig:Mrk421}
\end{figure}

A great opportunity for testing our speculation is available with the current and upcoming VHE telescopes. The Large High Altitude Air Shower Observatory (LHAASO, \citealt{LHAASO_WhitePaper_2019}) is sensitive in the 0.1 TeV to 1 PeV energy band, and the upcoming Cherenkov Telescope Array (CTA, \citealt{CTA_2011}) is sensitive in the 20 GeV to 300 TeV energy band\footnote{CTA official website: \url{https://www.cta-observatory.org/}}. Their synergy observations provide an unprecedented chance to investigate the particle acceleration and radiation physics of the TeV-PeV $\gamma$-rays. We show the delectability of the TeV excess in the SEDs for the four BL Lacs with the LHAASO and CTA in Fig.\ref{fig:instru&seds}, where one-year sensitivities are used. One can find that the TeV excess is not clearly shown up in the observed SEDs due to the strong EBL absorption effect, which usually flattens the observed spectrum above the 0.1 TeV range. In contrast, it should be confidently detectable with the CTA up to $\sim 10$ TeV for the four BL Lacs. The TeV excess in 1ES 0229+200 may also be marginally detectable with the LHAASO in the energy range of several TeV. Meanwhile, The TeV excess of Mrk 421 predicted by our model is detectable with the CTA and LHAASO up to $\sim 40$ TeV. Its spectrum in the $8 \sim 17$ TeV energy band is predicted as $F_\nu \propto \nu^{-1.8}$. We have proposed to verify this TeV excess using the first operation year data of the LHAASO.

The keV-MeV excesses of the four BL Lacs are around $0.1 \sim 10$ MeV. Although they are marginally in the energy range of the EGRET ($0.2 \sim 100$ MeV; \citealt{HartmanRC...EGRET1999}), it is not sensitive enough for the detection. However, the keV-MeV excess of H2356--309 was tentatively observed in the X-ray band, showing up like an X-ray plateau with a distinct feature of spectral hardening. Missions have been proposed for observations on the MeV $\gamma$-rays that have sensitivities higher than current/past missions by one or two orders of magnitude. For instance, the \textit{e-ASTROGAM} is designed to improve the instrument sensitivity in the energy range of $0.3 \sim 100$ MeV with one-year sensitivity of $\sim 1.3\times10^{-12}\,{\rm erg\,cm^{-2}\,s^{-1}}$ at $\sim 0.5$ MeV \citep{DeAngelisA...2018}. As shown in Fig.\ref{fig:instru&seds}, the keV-MeV excess should be detectable with the \textit{e-ASTROGAM}.


\section{Summary}
\label{sec:conc}
We have investigated the hadronic origin of the observed TeV excess for four selected TeV BL Lacs (1ES 0229+200, 1ES 0347--121, 1ES 1101--232, and H2356--309) with a single-zone leptohadronic model by considering the cascade emission within their jets in detail. We summarize our results as follows.
\begin{itemize}
    \item Their broadband SEDs are well represented by our model. The SED in the radio-optical-X-ray-GeV gamma-ray energy band is mainly attributed to the synchrotron radiations and the SSC process of the primary electron population. The model parameters are consistent with those in leptonic models. The TeV excess is explained with the VHE $\gamma$-ray emission from the p$\gamma$ process via the $\pi^{0}$ decay, assuming that the power-law index of the proton distribution is -2.2. The target photon fields are from the synchrotron radiations of the primary electron population and the cascade electron population produced via the internal $\gamma\gamma$ absorption and the p$\gamma$ process.

    \item The cascade synchrotron radiations result in an excess in the keV-MeV band of the SEDs, illustrated as a distinct bump or plateau as shown in the SEDs of 1ES 0347--121, 1ES 1101--232, and H2356--309. The keV-MeV excess enhances the production of VHE photons in the $p\gamma$ process and reduces the proton power by about one order of magnitude. However, the derived powers are still 3-4 orders of magnitude larger than the Eddington luminosity. Further tests by setting the power-law index of the proton energy distribution as -1.5 and enlarging the low energy boundary of the proton energy distribution, the derived proton power is reduced by a factor around 2-4. The breaking Eddington limit issue is still cannot be overcome. Thus, our model is challenged by the current accretion picture of AGNs.

    \item Tentative TeV excess and MeV excess were observed in nearby bright TeV source Mrk 421. As predicted by our model, its TeV excess is mainly contributed by the cascade p$\gamma$ emission, which dominates the observed SED beyond 5 TeV and can be detectable with the CTA and LHAASO up to $\sim 40$ TeV. The convex break at 235 MeV in the SED of Mrk 421 observed with the EGRET might be resulted from its MeV excess. Mrk 421 is the best candidate for testing our model by synergic observations of the TeV excess and keV-MeV excess with LHAASO, CTA, and future missions that are sensitive in the keV-MeV energy band.
\end{itemize}

\newpage
\begin{acknowledgements}
We thank Jin Zhang, Xiang Yu Wang, Ruo Yu Liu, and Rui Xue for help discussion. We also thank Jin Zhang for providing us the SED data. This work is supported by the National Natural Science Foundation of China (Grant No.12133003 and U1731239), Guangxi Science Foundation (grant No. 2017AD22006) and Innovation Project of Guangxi Graduate Education (YCBZ2021025).
\end{acknowledgements}

\begin{table}
\begin{center}
  	\caption{Leptohadronic Model Parameters}
	\label{tab:paras}
	\begin{tabular}{lcccccccccccccc}
    \hline\hline
    Source & z &  $\delta_{\rm D}$ & B & $\Delta t$ & $\gamma_{\rm e,min}$ & $\gamma_{\rm e, b}$ & $p_{1}$ & $p_{2}$ & $N_{0,e}$ & $E_{\rm p,max}$ & $E_{\rm p,cut}$  & $N_{\rm 0,p}$ \\
    ~ & ~ & ~ & $(\rm G)$ & $\rm hr$ & ~  & ~ & ~ & ~ & $(\rm cm^{-3})$ & $(\rm eV)$ & $(\rm eV)$ & $(\rm cm^{-3}\,eV^{-1} )$   \\
    \hline
    1ES 0229+200 & $0.139^{(1)}$ & 8.4 & 0.48 & 24 & 2 & $3.9 \times 10^{5}$ & 2.08 & 3.16 & $1.3 \times 10^{3}$ & $1 \times 10^{15}$ & $2.8 \times 10^{13}$ & $1.5 \times 10^{18}$\\
    \hline
    1ES 0347--121 & $0.188^{(2)}$ & 11 & 0.65 & 12 & 100 & $1.4 \times 10^{5}$ & 2.42 & 3.5 & $8.2 \times 10^{4}$ & $1 \times 10^{14}$ & $8 \times 10^{12}$ & $4.8 \times 10^{19}$\\
    \hline
    1ES 1101--232 & $0.186^{(3)}$ & 12 & 1.05 & 12 & 5 & $6.5 \times 10^{4}$ & 1.8 & 4.1 & 85 & $5 \times 10^{14}$ & $2.4 \times 10^{13}$ & $2.2 \times 10^{19}$\\
    \hline
    H2356--309 & $0.165^{(4)}$ & 7.6 & 0.5 & 24 & 2 & $6.3 \times 10^{4}$ & 2.1 & 3.4 & $3.8 \times 10^{3}$ & $1 \times 10^{15}$ & $2.8 \times 10^{13}$ & $1.1 \times 10^{19}$\\
    \hline
    Mrk 421 & $0.031^{(5)}$ & 29 & 0.058 & 3 & 20 & $4.5 \times 10^{5}$ & 2.36 & 4.0 & $1.9 \times 10^{5}$ & $1 \times 10^{15}$ & $2.8 \times 10^{13}$ & $4.5 \times 10^{19}$\\
    \hline
    \end{tabular}
\end{center}
\text{\textbf{References.}(1)\cite{Aharonian...2000};(2)\cite{Woo...2005};(3)\cite{Wolter...2000};(4)\cite{Bersanelli...1992};(5)\cite{Punch...1992}}
\end{table}

\clearpage

\clearpage

\clearpage

\clearpage

\bibliographystyle{aasjournal}

\end{document}